\documentclass[prb,twocolumn,amsmath,amssymb,aps,floatfix,superscriptaddress]{revtex4-2}
\usepackage{graphicx}
\usepackage{amssymb}
\usepackage{amsmath}
\usepackage{physics}
\usepackage{xcolor}
\usepackage[normalem]{ulem}
\usepackage{bm}
\usepackage{hyperref}
\usepackage{mathrsfs}
\usepackage{lipsum}
\newcommand{\rmm}[1]{{\textcolor{black}{#1}}}
\newcommand{\ccss}[1]{{\textcolor{black}{#1}}}
\begin{document}


\title{Stability of (Active) Bilayer Skyrmions in Synthetic Antiferromagnets
}

\author{Raí M. Menezes}
\email{rai.menezes@ufpe.br}
\affiliation{Departamento de F\'isica, Universidade Federal de Pernambuco, Cidade Universit\'aria, 50670-901, Recife-PE, Brazil}
\affiliation{Department of Physics, University of Antwerp, Groenenborgerlaan 171, B-2020 Antwerp, Belgium
}
\author{Clécio C. de Souza Silva}
\email{clecio.cssilva@ufpe.br}
\affiliation{Departamento de F\'isica, Universidade Federal de Pernambuco, Cidade Universit\'aria, 50670-901, Recife-PE, Brazil}

\begin{abstract}

Synthetic antiferromagnetic (SAF) skyrmions are nanoscale composite textures that exhibit high-speed, Hall-free current-driven motion and recently demonstrated self-propulsion. These remarkable properties rely on the stability of the SAF skyrmion's topological bound state, whose underlying mechanisms remain unclear. Here, using an atomistic spin model, we analyze the collapse pathways of bilayer SAF skyrmions in homochiral systems, where both ferromagnetic layers share the same Dzyaloshinskii–Moriya interaction (DMI) vectors, and in heterochiral systems, where the DMI vectors have opposite directions. We find that pair destruction occurs either by decoupling or by sequential collapse into the homogeneous antiferromagnetic state, so the activation energy is set by the smaller of these two barriers. By examining how these barriers vary with DMI strength, anisotropy, magnetic field, and interlayer exchange, we identify regimes of enhanced stability. In particular, increasing interlayer coupling strengthens homochiral skyrmions but weakens heterochiral ones, while reducing the anisotropy constant effectively stabilizes heterochiral SAF skyrmions. These results outline viable strategies to optimize SAF heterostructures for enhanced skyrmion stability in racetrack devices and emerging active skyrmionic systems.

\end{abstract}

\maketitle

\vskip 2pc

\section{Introduction}


The search for nanoscale information carriers capable of fast, energy-efficient motion has motivated extensive interest in magnetic skyrmions and related topological excitations. These spin configurations emerge in systems where exchange, anisotropy, and chiral interactions, such as the Dzyaloshinskii-Moriya interaction (DMI), compete~\cite{Roessler2006,nagaosa2013review}. In ultrathin ferromagnetic films interfaced with heavy metal layers, the DMI can be engineered by the interfacial spin-orbit coupling and broken inversion symmetry, enabling robust skyrmion formation even at room temperature and zero magnetic field~\cite{romming2013writing,Ma2016,Tacchi2017,Pollard2017,Brandao2019,Carvalho2023}. The small size and high topological stability of skyrmions in these materials make them suitable candidates for racetrack devices, logic architectures, and non-conventional computing~\cite{Tomasello2014,zhao2024electrical,luo2021skyrmion,yan2021,song2020skyrmion,raab2022brownian,beneke2024gesture}. Furthermore, compared with domain walls and other magnetic textures, skyrmions can be manipulated with considerably lower current densities, reducing Joule heating and facilitating device miniaturization~\cite{sampaio2013nucleation,woo2016observation}.

In ferromagnets, current-driven skyrmions experience a lateral gyrotropic force, proportional to their topological charge, that deflects their trajectory~\cite{DirectObsSkHE}. This phenomenon, known as the skyrmion Hall effect (SkHE), poses serious challenges for device reliability. Antiferromagnetic skyrmions have been proposed as an effective solution for this problem, as their compensated spin configuration leads to neutral total topological charge and negligible SkHE~\cite{Barker2016, zhang2016antiferromagnetic, Gobel2017}. Synthetic antiferromagnets (SAF), made of ultrathin ferromagnetic layers interacting via antiferromagnetic (AFM) Ruderman–Kittel–Kasuya–Yoshida (RKKY) coupling, offer a particularly attractive platform: they combine the SkHE-free dynamics of antiferromagnets with the high tunability of multilayer systems~\cite{zhang2016magnetic,KoshibaeNagaosa_2017,dohi2019formation,legrand2020room,correia2024stability,Lee2025}. SAF skyrmions have been experimentally shown to reach speeds far exceeding those of ferromagnetic skyrmions~\cite{HighSpeedSK_Dynamics_SAF}. More recently, it has been predicted that certain configurations of magnetic multilayers can host SAF skyrmions capable of self-propulsion and self-orientation, exhibiting autonomous motion driven solely by their internal, time-dependent degrees of freedom~\cite{deSouzaSilva2025}.


The functionality of SAF skyrmions as reliable information carriers and their application as active matter systems ultimately depends on the robustness of the underlying bound topological state against thermal fluctuations, defects, and external forcing. Despite rapid progress in understanding their dynamics, the mechanisms governing the formation, persistence, and annihilation of SAF skyrmion pairs remain poorly understood. A systematic investigation of their collapse pathways and stability across a wide parameter range is therefore essential for refining material design and evaluating their robustness for practical applications. 

\begin{figure}[t!]
\centering
\includegraphics[width=\columnwidth]{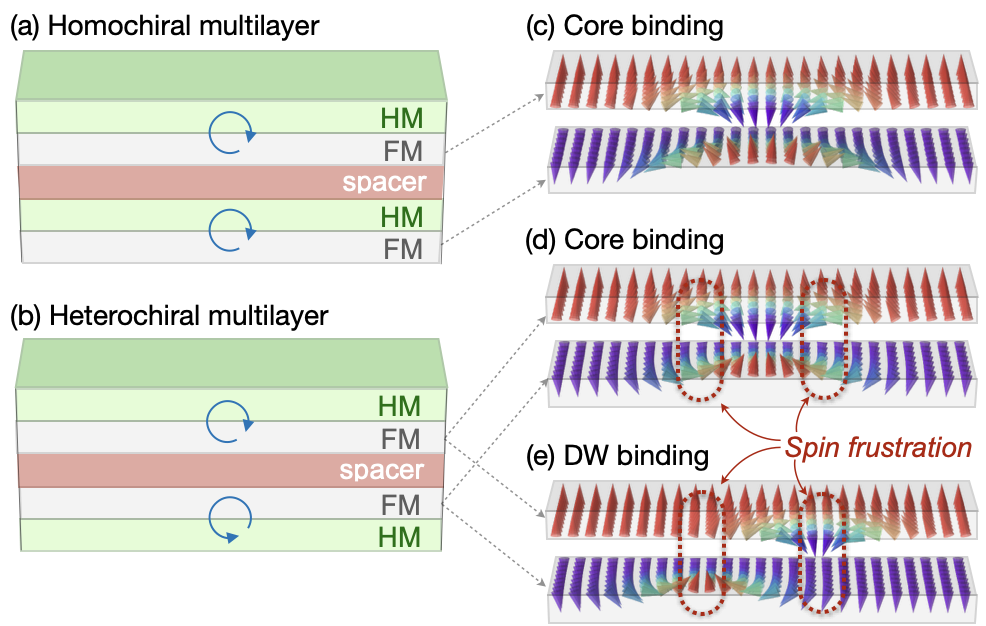}
\caption{ Schematic illustration of (a) homochiral and (b) heterochiral synthetic antiferromagnets, where the spin chirality of each FM layer (indicated by the circular arrows) is induced by interfacial DMI, which changes sign depending on the ferromagnet--heavy-metal (FM-HM) stacking order. (c)-(e) Cross-sectional view of the spin configuration for coaxial (core) binding in the (c) homochiral and (d) heterochiral cases, and for non-coaxial (domain-wall) binding (e).
}
    \label{fig.schematic}
\end{figure}

In this paper, we theoretically investigate the collapse mechanism of bilayer skyrmions in SAF and possible routes to enhance their stability in two configurations of interest: \emph{homochiral} and \emph{heterochiral} SAF multilayers, as illustrated in Fig.~\ref{fig.schematic}. In the homochiral configuration (a), both ferromagnetic layers share the same interfacial DMI vectors, favoring a perfect AFM alignment of the spins and a robust coaxial bound state between the skyrmions (c). In contrast, in heterochiral SAFs (b), the DMI vectors have opposite directions, leading to interlayer frustration and allowing both coaxial and non-coaxial binding, as shown in panels (d) and (e). These frustrated interactions generally weaken the binding compared with the homochiral case~\cite{correia2024stability}. At the same time, the finite bond length of the non-coaxial state provides internal degrees of freedom that enable self-propulsion~\cite{deSouzaSilva2025}, motivating a detailed analysis of their stability.

We find that, in both coaxial and noncoaxial configurations, the bound state can be destroyed either by decoupling the skyrmion pair or by the decay of the skyrmions into the homogeneous antiferromagnetic configuration of the layers. The latter case happens with the sequential collapse of skyrmions, each at its own time. Therefore, the activation energy for pair destruction can be taken as the smallest of the energy barriers for the collapse of the first skyrmion and for decoupling. We conduct a thorough investigation of these energy barriers as functions of several magnetic parameters, including DMI, anisotropy, magnetic field, and interlayer exchange, demonstrating that the stability of the pair can be significantly enhanced for appropriate parameter choices. In particular, we demonstrate that increasing the interlayer coupling improves the stability of homochiral skyrmions, while, in contrast, reducing the stability of heterochiral skyrmions. However, suitable combinations of an external magnetic field and unbalanced DMI constants, along with a reduction in the anisotropy constant, are shown to be effective strategies for preventing collapse and increasing the binding energy of heterochiral SAF skyrmions in the noncoaxial configuration.

The paper is organized as follows. In Sec.~\ref{secII}, we introduce the spin model employed to simulate the magnetic states and describe the numerical method used to compute minimum-energy paths for skyrmion collapse, which allows us to estimate their stability in the systems considered. Section~\ref{sec.III} presents our results. In Sec.~\ref{sec.III}A, we report the activation energies for the collapse of homochiral SAF skyrmions as functions of the DMI strength, magnetic anisotropy, and applied magnetic field, and we elucidate the specific collapse mechanism of these skyrmions. In Sec.~\ref{sec.III}B, we analyze the stability of heterochiral SAF skyrmions, demonstrating the existence of both coaxial and noncoaxial bound configurations and discussing the conditions under which each state is stabilized. Section~\ref{sec.III}C examines the activation energies for the collapse of heterochiral SAF skyrmions as functions of the relevant magnetic parameters, showing that although their stability is reduced compared to the homochiral SAF case, it can be significantly enhanced by appropriate parameter choices. Finally, in Sec.~\ref{sec.III}D, we compare the collapse and decoupling energies of SAF skyrmions, revealing that the destruction of heterochiral SAF skyrmions is more likely to occur via decoupling rather than collapse, and that the decoupling energy can also be tuned by adjusting the magnetic parameters. Section~\ref{sec.conclusion} concludes the paper.

\section{Theoretical framework}\label{secII}

\subsection{Spin model}\label{secIIA} 

To investigate the stability of skyrmions in bilayer synthetic antiferromagnets, we perform spin-dynamics simulations using the \textsc{Spirit} package~\cite{muller2019spirit}, modified to account for the specific magnetic interactions of our system. We consider a bilayer of classical spins in fcc stacking, with each layer arranged on a hexagonal lattice. The total magnetic energy is described by an extended Heisenberg Hamiltonian, 

\begin{equation}
    \mathcal{H}=\mathcal{H}^{(1)}+\mathcal{H}^{(2)}+\sum_i J_\text{int}\,\mathbf{n}_i^{(1)} \cdot \mathbf{n}_i^{(2)},
\end{equation}
where $\mathbf{n}_i^{(l)}$ denotes the orientation of the $i^\text{th}$ spin in layer $l$ (with $l=1$ or $2$), $J_\text{int}>0$ is the antiferromagnetic interlayer coupling constant, and $\mathcal{H}^{(l)}$ is the intralayer Hamiltonian of layer $l$, given by  

\begin{equation}
\begin{aligned}
  \mathcal{H}^{(l)}
    = & -\sum_{\langle i,j \rangle} \mathcal{J}^{(l)}\,\mathbf{n}_i^{(l)} \cdot \mathbf{n}_j^{(l)} 
        - \sum_{\langle i,j \rangle} \mathbf{D}_{ij}^{(l)} \cdot \big(\mathbf{n}_i^{(l)} \times \mathbf{n}_j^{(l)}\big) \\
      & - \sum_i K^{(l)} \,\big(\mathbf{n}_i^{(l)}\cdot\hat{z}\big)^2
        - \sum_i \mu \,\mathbf{B} \cdot \mathbf{n}_i^{(l)}.
    \label{HeisenbergHam}
\end{aligned}
\end{equation}
Here, $\mathcal{J}$ is the intralayer exchange interaction; $\mathbf{D}_{ij}=D(\hat{u}_{ij}\times\hat{z})$ the N\'eel-type Dzyaloshinskii--Moriya (DM) vector, with $\hat{u}_{ij}$ the unit vector between sites $i$ and $j$; $\mathbf{B}$ the applied magnetic field, $K$ the out-of-plane anisotropy constant, $\mu$ the magnitude of the magnetic moment, and $\langle i,j \rangle$ indicates nearest-neighbor (NN) spin pairs.  

The dynamics of the spin system are governed by the Landau--Lifshitz--Gilbert (LLG) equation,  

\begin{equation}
  \frac{\partial \mathbf{n}_i}{\partial t}
    = -\frac{\gamma}{(1+\alpha^2)\mu} \left[\mathbf{n}_i \times \mathbf{B}_i^\text{eff} + \alpha \,\mathbf{n}_i \times \big(\mathbf{n}_i \times \mathbf{B}_i^\text{eff}\big) \right],
\end{equation}
where $\gamma$ is the electron gyromagnetic ratio, $\alpha$ the Gilbert damping parameter, and $\mathbf{B}_i^\text{eff} = -\partial \mathcal{H}/\partial \mathbf{n}_i$ the effective field.  

For the simulations, we consider reference intralayer parameters for both layers based on Co/Pt thin films~\cite{rohart2016path, stosic2017paths}. We set $\mathcal{J}=29$~meV per bond, $D=1.5$~meV per bond, $K=0.279$~meV per atom, and $\mu = 2.1\mu_B$, where $\mu_B$ is the Bohr magneton. The site-to-site distance is $a = 2.51~\text{\AA}$. Variations in these quantities, as well as in the interlayer exchange $J_\text{int}$, will be explored throughout the manuscript. Unless stated otherwise, the parameters are set to these reference values.

\subsection{Minimum energy paths for skyrmion collapse}\label{GNEB}

To characterize the stability of skyrmions in our system, it is essential to identify the underlying collapse mechanism of such structures and the associated energy barriers. For this purpose, we compute the minimum energy paths (MEPs) of skyrmion collapse using the geodesic nudged elastic band (GNEB) method~\cite{bessarab2015method}, combined with the climbing image (CI) technique~\cite{henkelman2000improved}, both implemented in \textsc{Spirit}~\cite{muller2019spirit}. These methods allow us to determine the saddle point accurately ---the highest-energy configuration along the MEP---thereby providing the activation energy for the skyrmion destruction.  

In the GNEB approach, the MEP is represented by a discrete chain of $N_I$ magnetic configurations, or ``images,'' connecting the initial and final magnetic states. The initial guess of the path is given by the set $[\bm{\mathcal{M}}^1, \dots, \bm{\mathcal{M}}^{N_I}]$, where $\bm{\mathcal{M}}^\nu=(\mathbf{n}_1^\nu,\mathbf{n}_2^\nu,\dots,\mathbf{n}_N^\nu)$ denotes the $\nu^\text{th}$ configuration of a system with $N$ spins. In our calculations, these initial images are generated by applying a homogeneous rotation of the magnetic moments between the two target states, namely an initial state containing a single SAF skyrmion and a final homogeneous saturated SAF state.  

To relax this initial path toward the nearest MEP, the effective force acting on each image is computed from the negative energy gradient, $-\nabla \mathcal{H}^{\nu}$, where $\mathcal{H}^{\nu}$ is the energy of the $\nu^\text{th}$ configuration and $\nabla_i=\partial/\partial \mathbf{n}_i$. The component of this force along the tangent to the path is supplemented by an artificial spring force, ensuring a uniform distribution of images along the path. Meanwhile, the transverse component of the energy gradient drives the images toward the minimum-energy trajectory. The first and last images are fixed at the local minima corresponding to the initial and final states. Within the CI method, the spring forces acting on the highest-energy image are suppressed, while the force component along the path is inverted, causing this image to move uphill in energy until it converges to the saddle point. After the CI--GNEB calculation converges, the CI position coincides with the saddle point along the MEP, and the activation energy is obtained as the energy difference between this saddle point and the initial state.


\section{Results and discussion}\label{sec.III}

\subsection{Collapse of homochiral $\text{SAF}$ skyrmions}

\begin{figure}[t!]
\centering
\includegraphics[width=1.0\columnwidth]{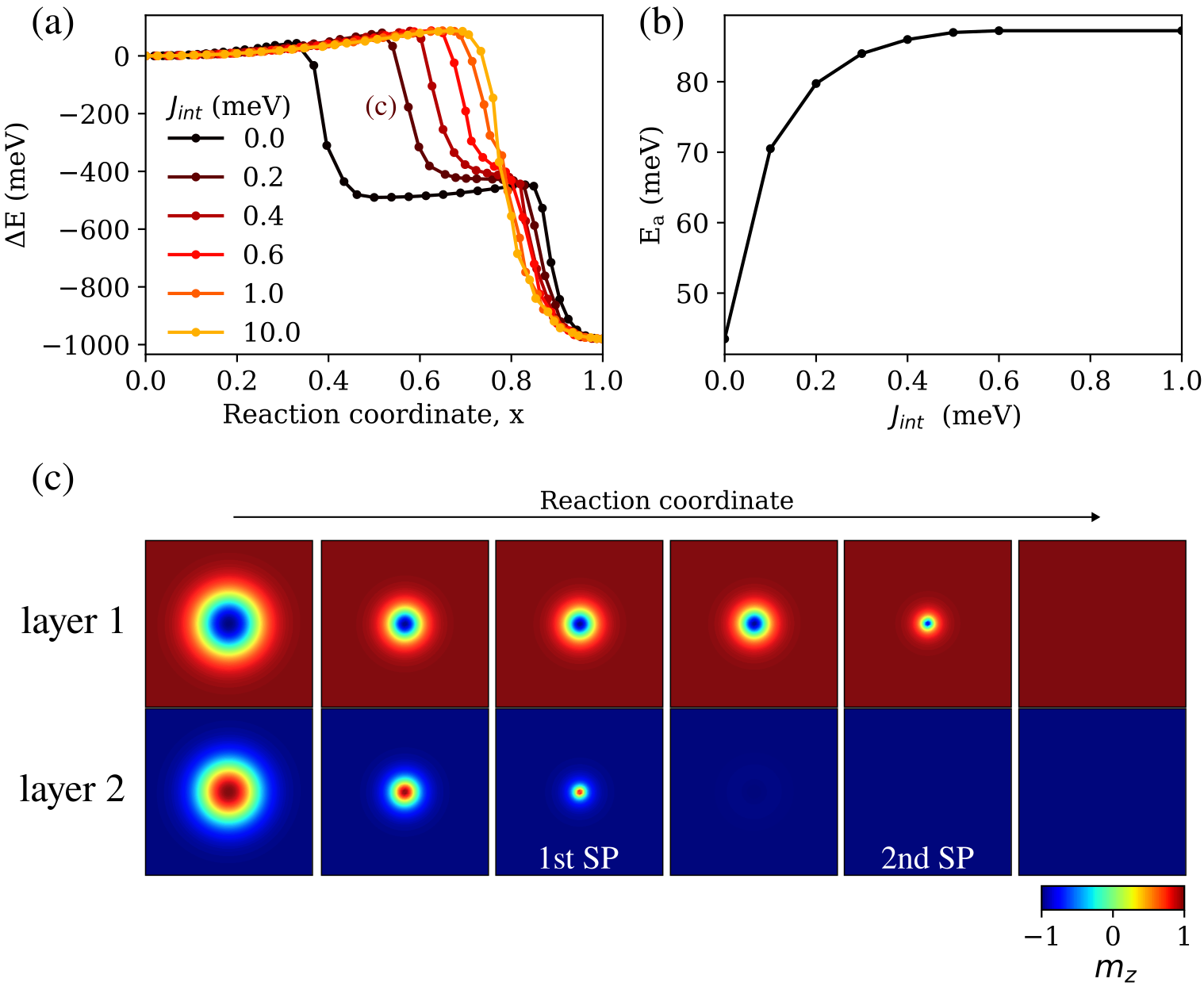}
\caption{ (a) MEPs for the collapse of an isolated SAF skyrmion for different values of the interlayer exchange coupling, $J_\text{int}$. The reaction coordinate, $x$, defines the normalized (geodesic) displacement along the collapse path, with $x=0$ representing the SAF skyrmion state and $x=1$ the homogeneous saturated SAF state. (b) Activation energy, $E_a$, for the collapse of the first skyrmion as a function of $J_\text{int}$. (c) Snapshots of the magnetic configurations at each layer along the MEP, for $J_\text{int}=0.2$~meV. The fist and second saddle-point (SP) configurations are indicated. 
}
    \label{fig1}
\end{figure}

To study the stability of SAF skyrmions, we perform MEP calculations, as described in Sec.~\ref{GNEB}, for the collapse of an isolated SAF skyrmion, which consists of a pair of skyrmions, one at each magnetic layer, with opposite topological charges and arranged in a coaxial configuration. Such configuration can be stabilized in homochiral layers, i.e., when the magnetic layers have same DMI chirality, as illustrated in \ccss{Fig.~\ref{fig.schematic}-(a) and (c)}, and as observed experimentally in Refs.~\onlinecite{legrand2020room,juge2022skyrmions}.   

Fig.~\ref{fig1}~(a) shows the calculated MEP for the collapse of an isolated SAF skyrmion for different values of the interlayer exchange coupling, $J_\text{int}$, where the intralayer magnetic parameters for both layers are set as those mentioned in Sec.~\ref{secIIA}, and at zero applied magnetic field. 

An interesting aspect of this collapse mechanism is the competition between the interlayer exchange coupling, which favors antiparallel alignment of the two layers and thus a symmetric collapse of the skyrmions in both layers, and the energy cost associated with the formation of Bloch points during skyrmion collapse, which instead favors independent collapse of the layers. As is well known for skyrmion annihilation, the collapse is characterized by the emergence of a Bloch point, which corresponds to the highest-energy configuration (saddle point) along the MEP. In three-dimensional systems, such as bulk films, the collapse typically proceeds layer by layer, thereby minimizing the energy cost along the MEP by allowing only one Bloch point to form at a time, rather than a homogeneous collapse with the simultaneous formation of multiple Bloch points~\cite{leishman2020topological}.

When the interlayer coupling is negligible, the MEP for the collapse of the SAF skyrmion exhibits two peaks (two saddle points), as shown by the black line in Fig.~\ref{fig1}~(a), indicating that the skyrmions collapse one layer at a time. In contrast, when the interlayer coupling is strong, the AFM alignment is enforced and the skyrmions in both layers collapse simultaneously, resulting in an MEP with a single peak (single saddle point), as shown by the yellow curve in Fig.~\ref{fig1}~(a).

At intermediate values of $J_\text{int}$, the competition between these two effects gives rise to the following collapse mechanism [see Fig.~\ref{fig1}~(c)]: (\textit{i}) both skyrmions initially shrink together, thereby minimizing the interlayer exchange energy; (\textit{ii}) once they reach a threshold radius—where the exchange energy difference associated with collapsing a single layer becomes smaller than the energy cost of forming a saddle point—one skyrmion collapses while the other remains at nearly the same size, bringing the MEP to its first saddle point; (\textit{iii}) after the first skyrmion collapses, the second skyrmion continues to shrink until it collapses as well, leading to the formation of another Bloch point and a second saddle point along the MEP.

Since the SAF skyrmion in our system is defined as a pair of skyrmions, one in each layer, we assess the stability of this configuration by calculating the activation energy, \(E_a\), required for the collapse of the skyrmion in at least one of the layers. 
Figure~\ref{fig1}(b) shows the calculated \(E_a\) for the collapse of the first skyrmion, defined as the energy difference between the first saddle point along the MEP and the initial SAF skyrmion state, as a function of the interlayer coupling strength. 
It is observed that the activation energy increases rapidly with \(J_\text{int}\), rising from \(E_a = 42~\text{meV}\) \rmm{($\approx 487 $~K)} at \(J_\text{int} = 0\) to \(E_a = 88~\text{meV}\) \rmm{($\approx 1021 $~K)} for \(J_\text{int} > 0.6~\text{meV}\), where it saturates.

For larger $J_\text{int}$ values, the spin configurations of the two layers are exactly antiparallel in both the initial state and the saddle point. Consequently, there is no longer any interlayer exchange energy difference between the two states, and the activation energy remains constant, with the energy cost arising solely from intralayer contributions.

\begin{figure}[t]
\centering
\includegraphics[width=1.0\columnwidth]{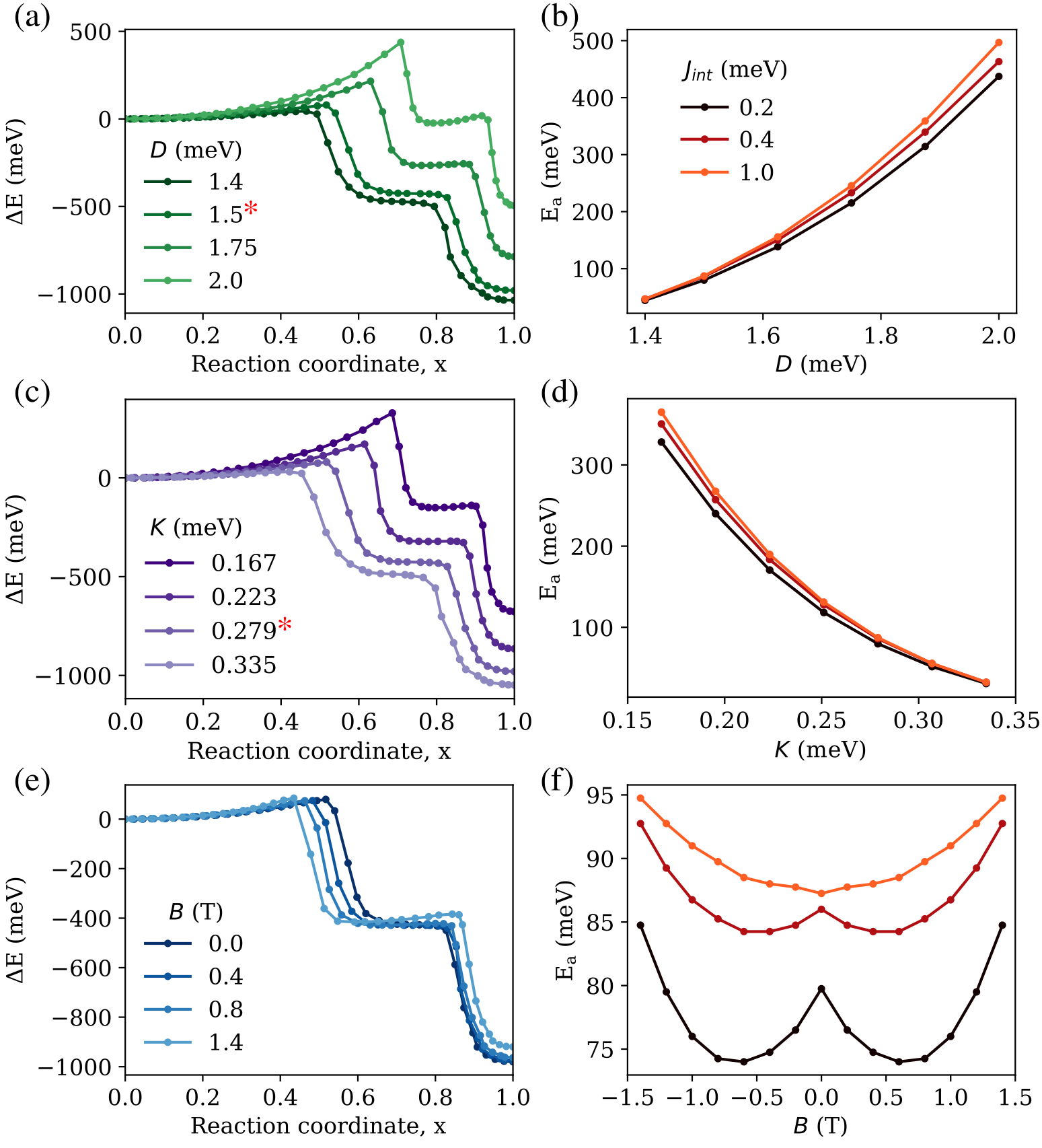}
\caption{ (a,c,e) MEPs for the collapse of a SAF skyrmion, calculated for $J_\text{int}=0.2$~meV and different values of the DMI strength $D$ (a), anisotropy constant $K$ (c), and applied field $B$ (e). Reference values are indicated by the red asterisk. (b,d,f) Activation energy $E_a$ for the collapse of the first skyrmion as a function of $D$ (b), $K$ (d), and $B$ (f), calculated for $J_\text{int}=0.2$, $0.4$, and $1.0$~meV.
}
    \label{fig2}
\end{figure}

We now explore how variations in the magnetic parameters affect the activation energy, i.e., the stability of the SAF skyrmion. For this purpose, we select three representative values of interlayer interaction, specifically $J_\text{int}=0.2, 0.4$, and $1.0$~meV, and examine the influence of DMI, anisotropy, and applied field strength for each case. Fig.~\ref{fig2}~(a,c,e) illustrates the MEPs for the collapse of a SAF skyrmion, calculated for $J_\text{int}=0.2$~meV and different values of DMI [Fig.~\ref{fig2}~(a)], anisotropy [Fig.~\ref{fig2}~(c)], and applied field [Fig.~\ref{fig2}~(e)], while the other intralayer magnetic parameters are fixed as described in Sec.~\ref{secIIA}. 

Increasing the DMI strength $D$ (in both layers) strongly enhances the activation energy for the collapse of the first skyrmion, as reflected by the higher first peak along the MEP in Fig.~\ref{fig2}~(a). The dependence of $E_a$ on $D$ for the three selected values of $J_\text{int}$ is shown in Fig.~\ref{fig2}~(b). In all cases, we observe a pronounced increase of $E_a$ with $D$, with the activation energy reaching nearly five times its initial value when $D$ is increased from $1.5$ to $2.0$~meV.

The stability of the SAF skyrmion can also be strongly affected by the anisotropy parameter. As shown in Fig.~\ref{fig2}~(c), the activation energy for the collapse of the first skyrmion increases significantly when the anisotropy is reduced. The dependence of $E_a$ on $K$ for the three selected values of $J_\text{int}$ is presented in Fig.~\ref{fig2}~(d). The enhanced stability obtained by increasing DMI or reducing anisotropy is also observed for \rmm{isolated FM skyrmions~\cite{stosic2017paths,cortes2017thermal}} and can be directly linked to the corresponding change in skyrmion size: both parameter variations enlarge the skyrmion in each layer, thereby making it more stable.

On the other hand, the stability of the SAF skyrmion exhibits a nonmonotonic dependence on the applied magnetic field. Figure~\ref{fig2}(f) shows the variation of \(E_a\) as a function of \(B\). For small field values, the activation energy decreases relative to the zero-field case, but above a threshold field this trend reverses and \(E_a\) starts to increase. This nonmonotonic behavior reflects the presence of competing interactions. The applied field affects skyrmion size differently in the two layers: the skyrmion shrinks in the layer where the core magnetization is opposite to the field direction, whereas it expands in the layer where the core magnetization is aligned with the field. A reduction in size within a layer lowers the intralayer energy barrier to the collapse of the smaller skyrmion. However, as the field increases, the mismatch in skyrmion sizes between the two layers becomes larger, thereby increasing the energy cost associated with the interlayer coupling and causing \(E_a\) to rise with \(B\) at higher fields. For larger \(J_\text{int}\) values, the spin configurations of the two layers tend to be antiparallel, which reduces the size mismatch. For sufficiently strong \(J_\text{int}\), the curve tends to flatten, as the magnetization configurations in the two layers become exactly opposite, and the Zeeman energies of the two layers cancel each other out.

\subsection{
\ccss{Heterochiral SAF skyrmions: coaxial vs. noncoaxial binding}
}

\begin{figure}[t]
\centering
\includegraphics[width=1.0\columnwidth]{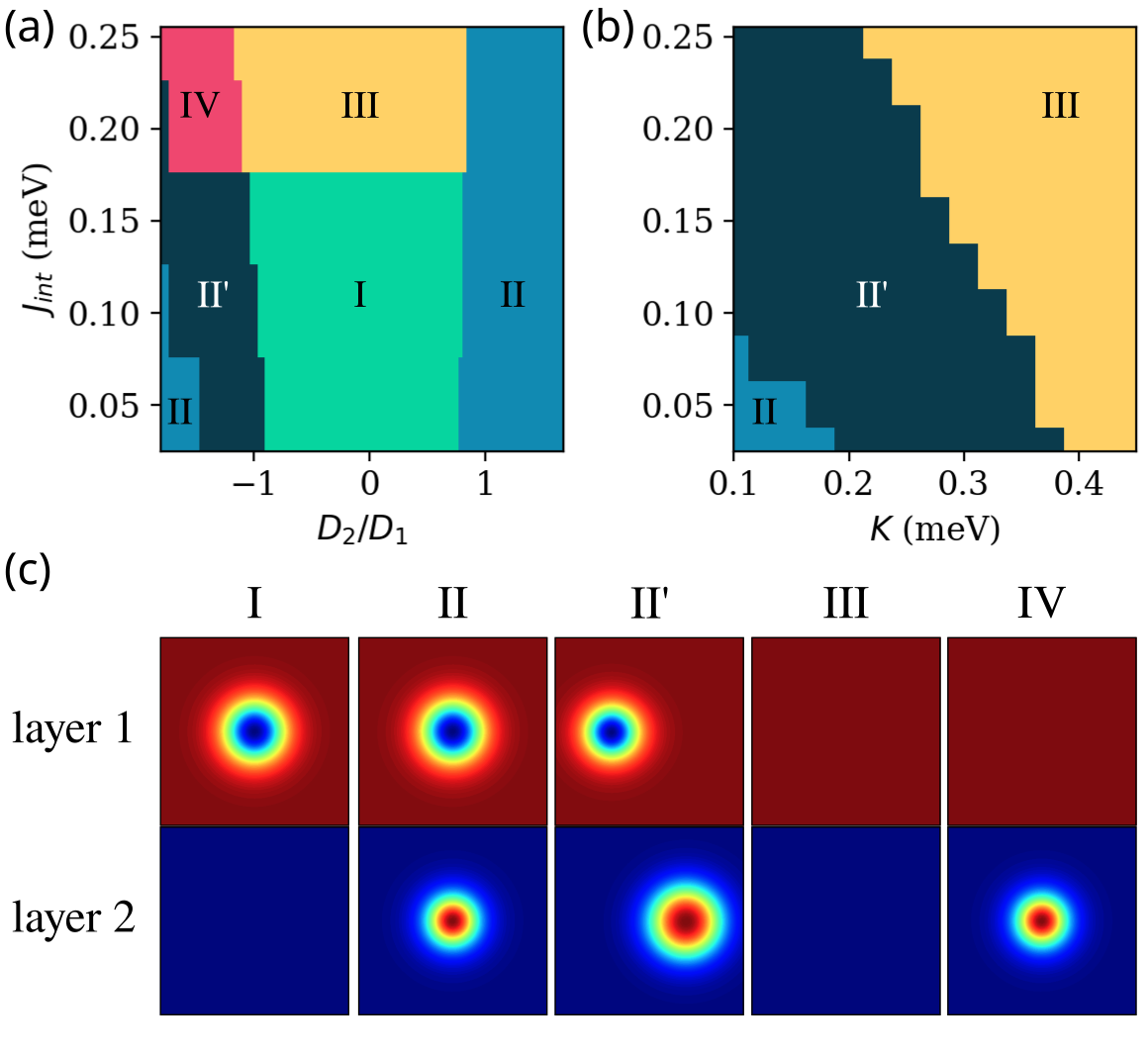}
\caption{
(a,b) Magnetic phase diagrams obtained from simulations for a bilayer SAF film as a function of the interlayer exchange coupling \( J_\text{int} \) and \rmm{(a) the DMI ratio $D_2/D_1$ between the two layers, for $D_1 = 1.5$~meV and $K=0.279$~meV, and (b) the anisotropy constant $K$ for the case $D_2 = -D_1 = -1.5$~meV.} (c) Corresponding snapshots of the stabilized magnetic configurations within the phase diagram for both layers of the sample. Colors represent the \( z \)-component of the magnetization, as in Fig.~\ref{fig1}(c).
 }
    \label{fig3}
\end{figure}

In contrast to the homochiral case, heterochiral SAF multilayers can stabilize two different configurations of bilayer skyrmion pairs: coaxial and noncoaxial~\cite{correia2024stability}. To map the conditions under which these configurations can be stabilized, we performed atomistic simulations as a function of interlayer coupling, perpendicular anisotropy, and the DMI ratio of the two layers, $D_2 / D_1$, where $D_1$ and $D_2$ are the DMI strengths of layers 1 and 2, respectively. The results are presented in the magnetic phase diagrams shown in Fig.~\ref{fig3}~(a,b). In the simulations, we use as the initial magnetic configuration a saturated SAF state with inverted magnetization in a small region corresponding to the skyrmion cores. The system is then relaxed via energy minimization, thereby stabilizing (meta)stable skyrmionic states.

Two distinct regions of the phase diagram support the formation of skyrmion pairs: region~II in Fig.~\ref{fig3}~(a,b), where conventional (coaxial) SAF skyrmions are stabilized; and, for \( D_2 / D_1 < 0 \), region~II$'$ in Fig.~\ref{fig3}~(a,b), where noncoaxial (NC)-SAF skyrmions appear. In both cases, the skyrmions may have different sizes in the two layers when \(|D_2/D_1|\neq 1\). The remaining regions of the phase diagram either do not stabilize skyrmions or stabilize them in only one layer.

 To better understand why heterochiral SAFs allow two different skyrmion bound states, we first notice that the opposite handedness of spin twists in both layers inevitably leads to some degree of frustration of the interlayer AFM interaction, even when the skyrmions have the same size ($|D_2/D_1|=1$). For the coaxial configuration, frustration occurs at the domain-wall region, whereas for noncoaxial binding, it occurs at the skyrmion cores, as illustrated in Fig.~\ref{fig.schematic}-(d) and (e). For the coaxial configuration, the spins at the skyrmion cores align antiferromagnetically, which reduces the total interlayer coupling energy $\mathcal{H}_\text{int}$ by roughly $E_\text{core}\approx \frac{2J_\text{int}}{\sqrt{3}a^2}\pi R^2$, where $R$ is the skyrmion radius. In contrast, at the domain wall region, of width $w$, the spins are essentially parallel to each other, which locally frustrates the AFM alignment and increases $\mathcal{H}_\text{int}$ by $E_\text{DW}\approx  \frac{2J_\text{int}}{\sqrt{3}a^2}\pi[(R+w/2)^2-(R-w/2)^2] =  \frac{2J_\text{int}}{\sqrt{3}a^2}\pi(2Rw)$. Therefore, one can define the threshold 
 \begin{equation}
    E_\text{core}/E_\text{DW} \approx 2R/w = 1,
 \end{equation}
 at which the destabilizing contribution of the spin frustration at the domain wall equals the stabilizing contribution of the cores. We argue that below this threshold, that is, for $R\lesssim2w$, the coaxial configuration becomes unstable. Of course, this is to be taken as a rough estimate, as in our analysis we have neglected contributions from other terms in the Hamiltonian, such as DMI and intralayer exchange, and we treat the domain wall as a uniform ring of width $w$. 

\begin{figure}[t]
\centering
\includegraphics[width=0.90\columnwidth]{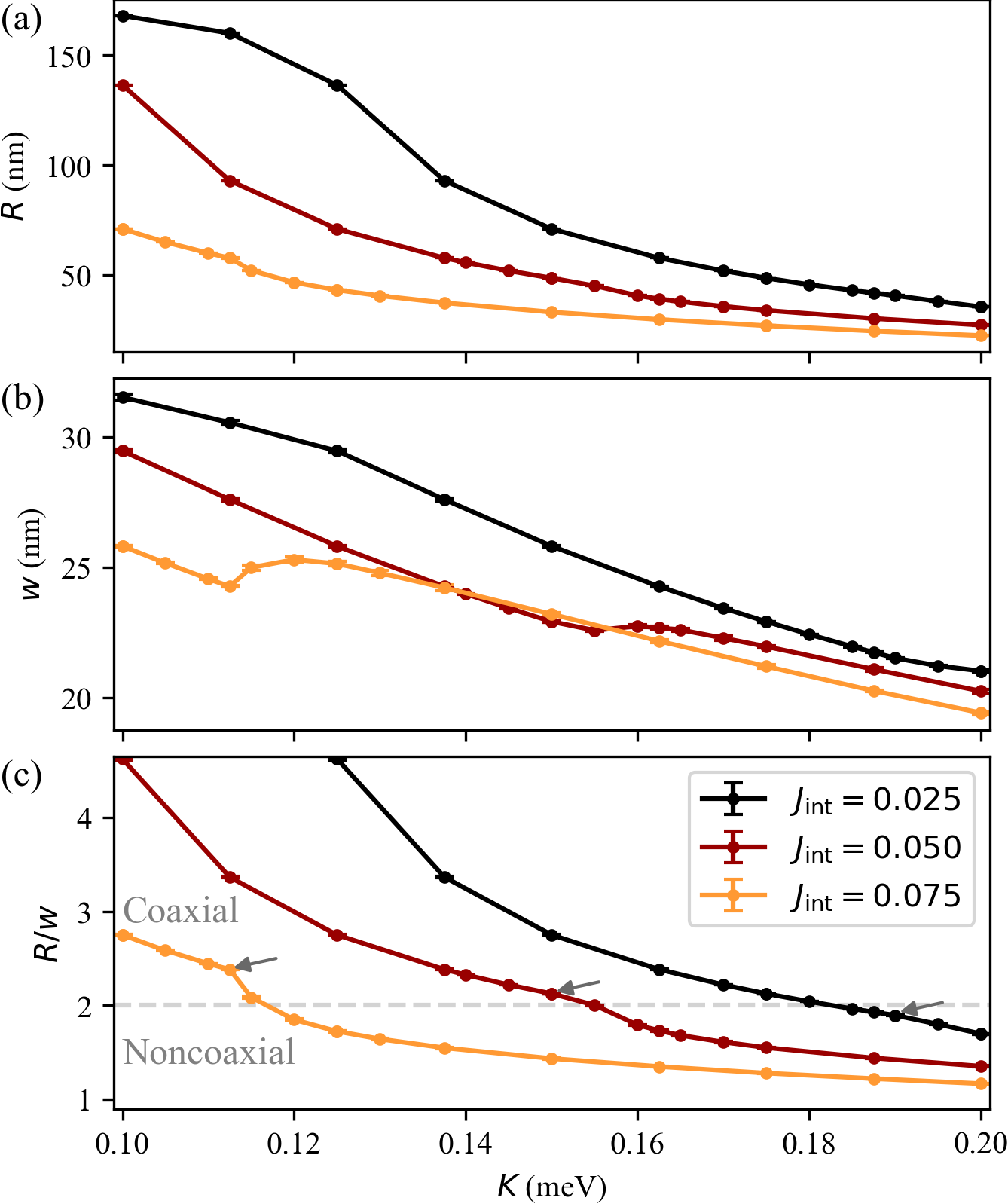}
    \caption{
      (a) Skyrmion radius $R$, (b) domain wall width $w$, and (c) $R/w$ ratio as a function of perpendicular anisotropy $K$ for a bilayer skyrmion in symmetric heterochiral SAFs with different interlayer exchange constants $J_\text{int}$ \rmm{and $D_2=-D_1=-1.5$ meV}. Error bars indicate standard errors of the fitting parameters $R$ and $w$ and propagated errors of $R/w$. The arrows in (c) indicate for each $J_\text{int}$ the upper limit of the coaxial arrangement, which is followed by the non-coaxial configuration. The dashed line indicates the $R=2w$ condition below which the estimated domain-wall frustration energy of the coaxial configuration exceeds the AFM core energy (see text). 
    }
\label{fig.CvsNC}
\end{figure}

 Fig.~\ref{fig.CvsNC} shows the radius $R$ and domain wall width $w$ of the skyrmion at layer 1 (same values as those of layer 2) for a symmetric ($|D_1|=|D_2|$) heterochiral SAF as a function of $J_\text{int}$ and $K$, calculated by fitting the skyrmion ansatz $m_z=\cos\theta_\text{fit}$, with $\theta_\text{fit} =2\arctan[\sinh(r/w)/\sinh(R/w)]$, to the atomistic simulation data. The transition between the coaxial and noncoaxial binding states occurs as $K$ increases, at the points indicated by the arrows \rmm{in Fig.~\ref{fig.CvsNC}~(c)}. Notice that the parameter window $0.1~\text{meV}\leq K\leq0.2~\text{meV}$ and $0.025~\text{meV}\leq J_\text{int}\leq0.075~\text{meV}$ spans a broad range of skyrmion morphologies, with $R$ varying from $\sim30$ nm to 170 nm and $R/w$ from $\sim1$ to 5. Still, the coaxial state becomes unstable with respect to the formation of a noncoaxial configuration at a value of $R/w$ close to the threshold line \rmm{($R=2w$)} estimated above. Therefore, despite the wide variation of $R$, $w$, and the details of the microscopic parameters, $R\lesssim2w$ can be taken as a quick-and-dirty rule of thumb for the stabilization of non-coaxial skyrmion-skyrmion binding in symmetric, heterochiral SAFs.



\subsection{Collapse of non-coaxial (NC) $\text{SAF}$ skyrmions}

In this section, we investigate the stability of skyrmions in heterochiral SAF films, particularly in the scenario where the skyrmions form a non-coaxial (NC) configuration [see Fig.~\ref{fig.schematic}-(e)], which is required for the aforementioned self-propulsion behavior. We therefore refer to this configuration as an NC-SAF skyrmion.

\begin{figure}[t]
\centering
\includegraphics[width=\columnwidth]{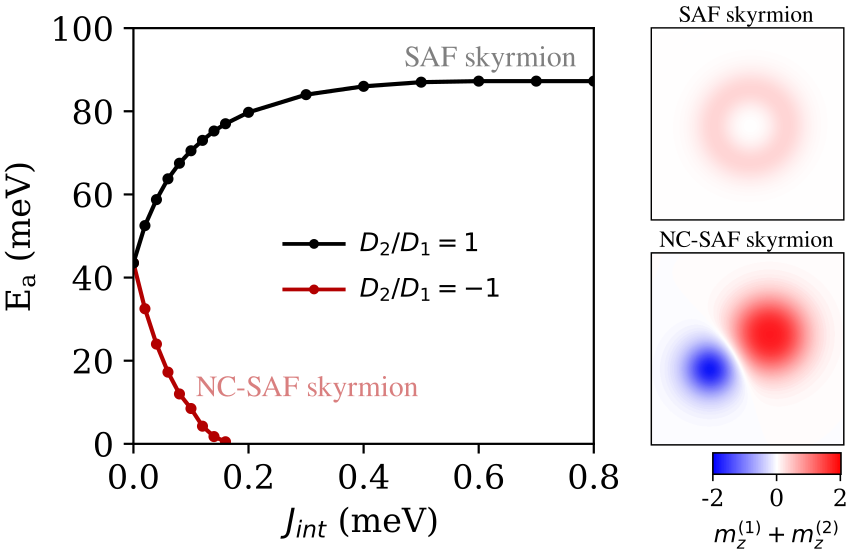}
\caption{
Calculated activation energy for skyrmion collapse as a function of the interlayer exchange coupling, for \( D_2 / D_1 = 1 \) (SAF skyrmion) and \( D_2 / D_1 = -1 \) (NC-SAF skyrmion). The insets on the right show the net magnetization of the bilayer system, i.e., \( m_z^{(1)} + m_z^{(2)} \), for both types of SAF skyrmions, in the case of \( J_\text{int} = 0.1~\text{meV} \). Note that the noncoaxial nature of the NC-SAF skyrmion is clearly revealed by its dipole-like configuration, exhibiting regions of positive and negative net magnetization. To illustrate the conventional SAF skyrmion in the inset, we choose \( D_2 \neq D_1 \) so that the skyrmions in the two layers have different sizes and the net magnetization is nonzero.
}
    \label{fig4}
\end{figure}

To compare the stability of NC-SAF skyrmions with that of conventional SAF skyrmions, we calculate the activation energy for their collapse as a function of the interlayer exchange coupling, for \( D_2 / D_1 = 1 \) (SAF skyrmion) and \( D_2 / D_1 = -1 \) (NC-SAF skyrmion). The results are shown in Fig.~\ref{fig4}. Here, we consider $|D_2|=|D_1|=1.5$~meV. Notice that, in contrast to conventional SAF skyrmions---whose stability increases with \( J_\text{int} \)---the stability of NC-SAF skyrmions decreases as the interlayer coupling increases. The NC-SAF skyrmion exhibits the highest activation energy at \( J_\text{int} = 0 \), but becomes unstable (i.e., \( E_a \to 0 \)) for \( J_\text{int} > 0.16~\text{meV} \), in agreement with the phase diagram in Fig.~\ref{fig3}~(a). Considering that a finite positive \( J_\text{int} \) is required to maintain the skyrmion pair bound in the bilayer system, preventing depairing during its dynamics (as required for applications), this result indicates that NC-SAF skyrmions are significantly less stable than both conventional SAF skyrmions and monolayer ferromagnetic skyrmions (the case \( J_\text{int} = 0 \)). In other words, although NC-SAF skyrmions have been proposed as promising candidates for current-free operation and active-matter behavior, under the considered parameters (derived from \rmm{Co/Pt systems with atom-thick magnetic layers}), they may be easily destabilized by thermal fluctuations or external perturbations, and therefore might not represent long-lived states suitable for \rmm{room-temperature} 
applications.

\rmm{Aiming to obtain NC-SAF skyrmions that are functional over a broader temperature range}, we now focus on methods to manipulate and enhance the activation energy for their collapse, thereby increasing their stability. For instance, this can be achieved by varying the relevant magnetic parameters of the SAF system, such as the DMI, anisotropy, and applied field.

\begin{figure}[t]
\centering
\includegraphics[width=1.0\columnwidth]{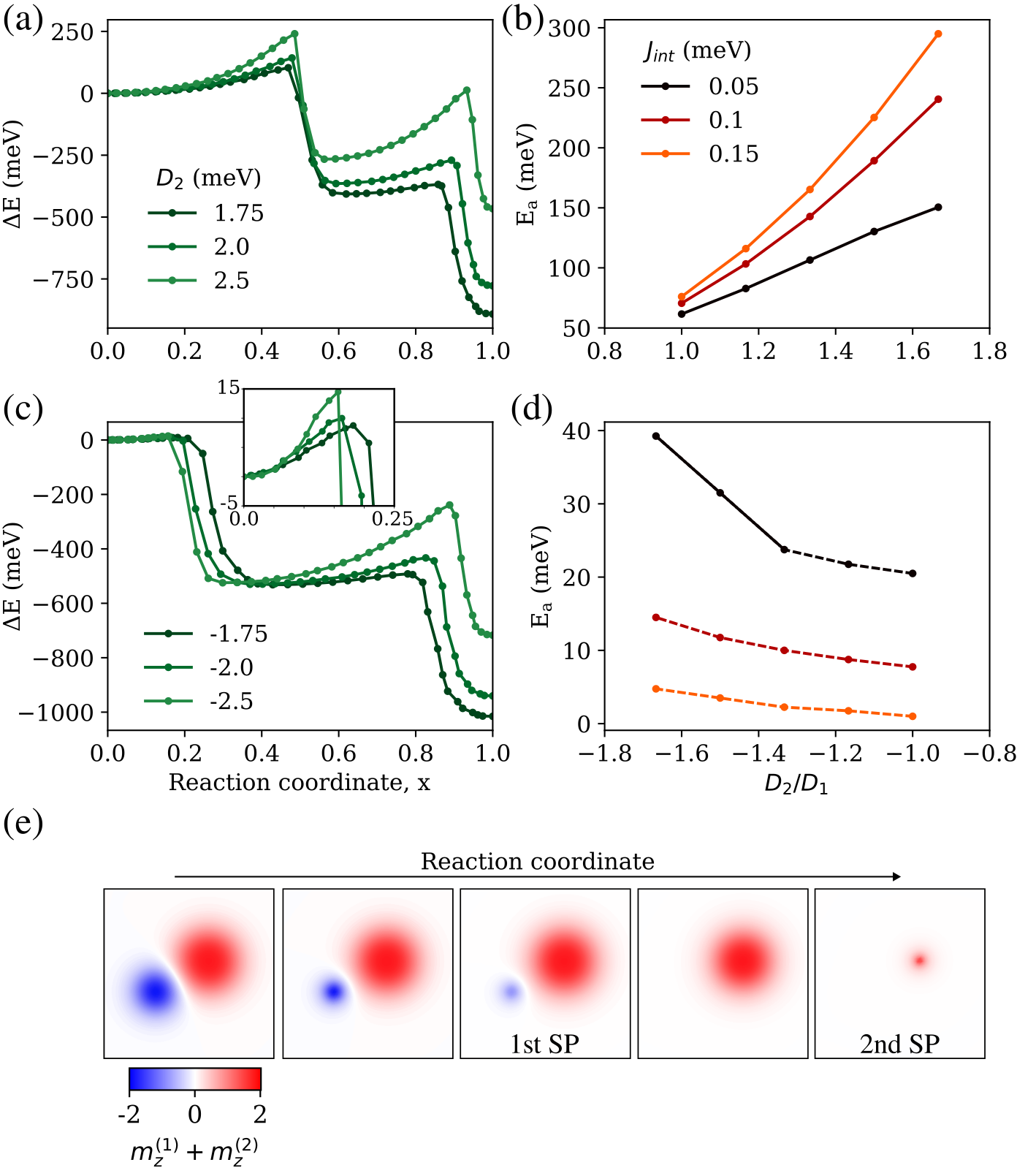}
\caption{
(a,c) MEPs for the collapse of (a) a SAF skyrmion and (c) a NC-SAF skyrmion for different values of the DMI strength \( D_2 \), with \( D_1 = 1.5~\text{meV} \) and \( J_{\text{int}} = 0.1~\text{meV} \). The inset in (c) shows a zoomed view of the vicinity of the first peak along the MEP. (b,d) Corresponding activation energies for the first saddle point (SP) of the SAF and NC-SAF skyrmions, respectively, for different \( J_{\text{int}} \) values. The dashed lines in (d) correspond to cases where NC-SAF skyrmions are stabilized, whereas the solid lines represent the coaxial configuration. (e) Snapshots of the net magnetization along the MEP for the collapse of the NC-SAF skyrmion for \( D_2 = 1.75~\text{meV} \), \( D_1 = 1.5~\text{meV} \), and \( J_{\text{int}} = 0.1~\text{meV} \).
}
    \label{fig5}
\end{figure}

Fig.~\ref{fig5}(a,c) shows the MEP for the collapse of (a) a SAF skyrmion and (c) a NC-SAF skyrmion for different values of the DMI strength \( D_2 \), while \( D_1 = 1.5~\text{meV} \) is kept fixed and \( J_\text{int} = 0.1~\text{meV} \). In the first case, increasing \( D_2 \) leads to a rapid rise of the energy barriers along the MEP; consequently, the activation energy for the collapse of the SAF skyrmion is strongly enhanced, as shown in Fig.~\ref{fig5}(b) for different values of \( J_\text{int} \). In contrast, for the NC-SAF skyrmion, although the energy barrier associated with the collapse of the second skyrmion (in layer 2, where the DMI is increased) is significantly enhanced, increasing \( |D_2| \) only slightly raises the activation energy for the collapse of the first skyrmion, as indicated by the inset in Fig.~\ref{fig5}(c) and shown in Fig.~\ref{fig5}(d) for different \( J_\text{int} \) values. 

The distinct response of the two states to the increase in \( |D_2| \) arises from the fact that, in the NC-SAF configuration, the skyrmion cores are not aligned. Therefore, increasing the skyrmion size---imposed by a larger DMI---does not favor the interlayer exchange energy. Instead, since the skyrmions are aligned through their domain walls, the interlayer exchange energy is minimized by increasing the domain-wall width rather than the skyrmion size. 

Fig.~\ref{fig5}(e) shows snapshots of the net magnetization along the MEP for the collapse of the NC-SAF skyrmion. The two saddle-point configurations are indicated and correspond to the collapse of the skyrmions in the first and second layers, respectively.

\begin{figure}[t]
\centering
\includegraphics[width=\linewidth]{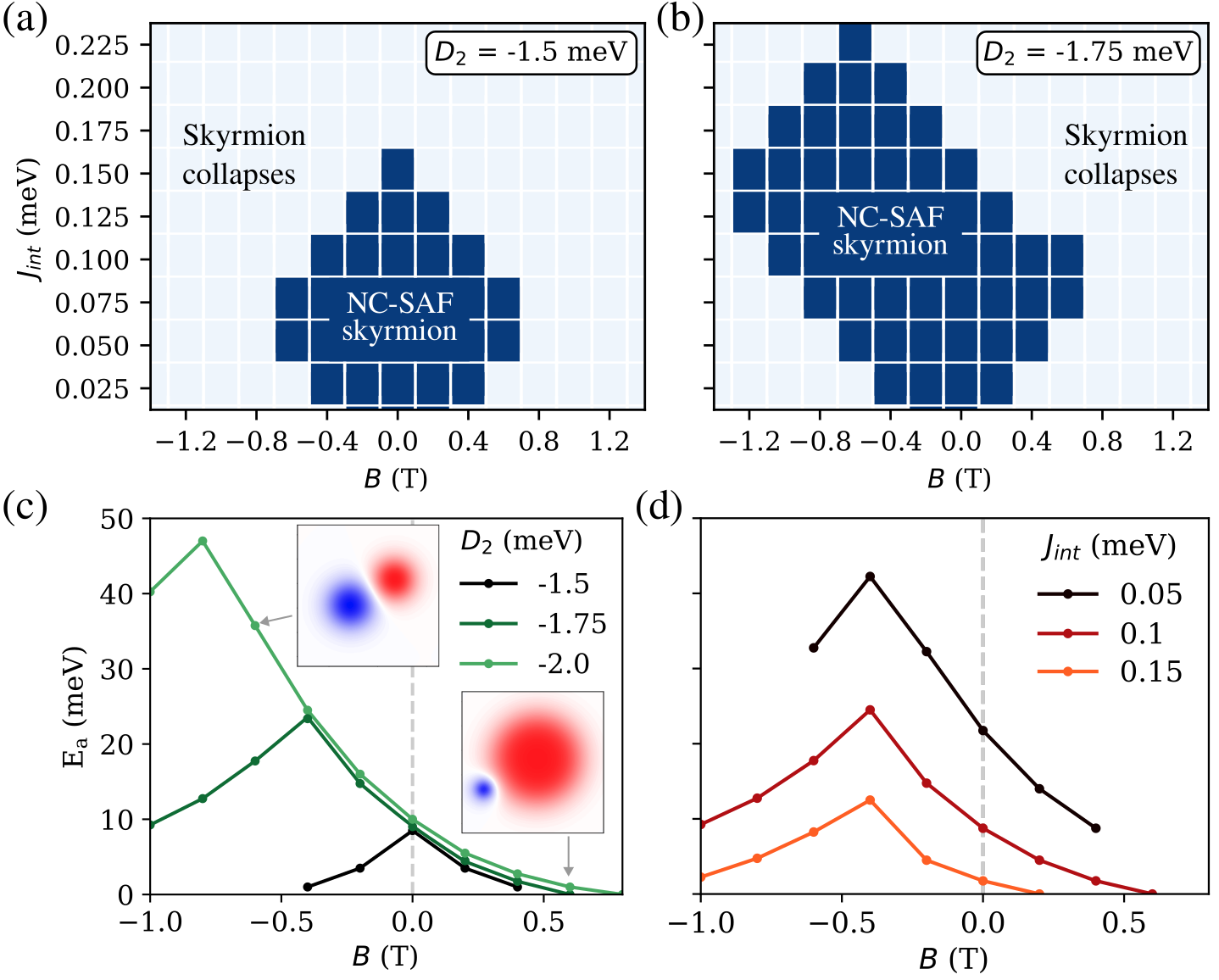}
\caption{
(a,b) Phase diagrams for skyrmion stability obtained from simulations as a function of \( J_\text{int} \) and \( B \), for \( D_1 = 1.5~\text{meV} \) with (a) \( D_2 = -1.5~\text{meV} \) and (b) \( D_2 = -1.75~\text{meV} \). The region where the NC-SAF skyrmion is stabilized is shown in dark blue. (c) Calculated activation energy profiles as a function of \( B \), for \( J_\text{int} = 0.1~\text{meV} \) and different values of the DMI strength \( D_2 \). The insets illustrate the variation in skyrmion size with the magnetic field for the case \( D_2 = -2.0~\text{meV} \). (d) Calculated activation energy profiles as a function of \( B \), for \( D_2 = -1.75~\text{meV} \) and different values of the interlayer exchange \( J_\text{int} \).
}
    \label{fig6}
\end{figure}

Let us now investigate the effect of an external magnetic field applied perpendicular to the film plane, i.e., $\textbf{B} = B\hat{z}$, on the stability of the NC-SAF skyrmion. Figure~\ref{fig6}(a) shows the magnetic phase diagram obtained from our simulations for the case $D_2/D_1 = -1$, as a function of $J_\text{int}$ and $B$. The region where the NC-SAF skyrmion can be stabilized is shown in dark blue. Note that, in this scenario, the phase diagram is symmetric with respect to positive and negative fields, since skyrmions in both layers have the same size and respond symmetrically to the field: the skyrmion core shrinks in the layer where its magnetization is opposite to the field direction and expands in the layer where it is aligned with the field. This behavior is in agreement with the phase diagram reported in Ref.~\cite{deSouzaSilva2025} for a similar SAF system.

In contrast, when $D_2/D_1<-1$, the phase diagram becomes asymmetric, as shown, for example, in Fig.~\ref{fig6}(b), where $D_2 = -1.75~\text{meV}$ and $D_1 = 1.5~\text{meV}$. In this case, skyrmions have different sizes at zero field. When the applied field causes the smaller skyrmion to shrink (positive $B$ in Fig.~\ref{fig6}(b)), the overall skyrmion stability is reduced. Conversely, when the field induces an expansion of the smaller skyrmion (negative $B$ in Fig.~\ref{fig6}(b)), the stability is enhanced, and the region of the phase diagram where the NC-SAF skyrmion can be stabilized extends toward negative values of $B$.

Figure~\ref{fig6}(c) shows the calculated activation energy profiles as a function of $B$, for $J_\text{int} = 0.1$~meV and different values of the DMI strength $D_2$. Note that when $|D_2| = |D_1|$ (the case $D_2 = -1.5$~meV in Fig.~\ref{fig6}(c)), the activation energy is symmetrically reduced for both positive and negative fields. In contrast, for cases where $|D_2| \neq |D_1|$, the activation energy can be largely enhanced for negative values of $B$. Since the skyrmions have different sizes at zero field, the first collapse along the MEP occurs in the layer where the skyrmion is smaller (i.e., where the DMI is weaker). By applying a field that expands this smaller skyrmion, its collapse becomes more difficult, thereby increasing the activation energy. The corresponding changes in skyrmion size are illustrated in the inset of Fig.~\ref{fig6}(c). When the field becomes sufficiently large, the skyrmion that was initially larger than the other becomes the smaller one. The collapse then occurs through that layer, and $E_a$ starts to decrease for larger field magnitudes, as this skyrmion shrinks. This explains the nonmonotonic behavior of $E_a$ for negative field values. Figure~\ref{fig6}(d) shows the energy profiles for different values of $J_\text{int}$, for $D_2 = -1.75$~meV, indicating that although $J_\text{int}$ affects the absolute value of the activation energy, the field at which $E_a$ reaches its maximum remains constant for a given DMI ratio, i.e., it depends only on the difference between the intralayer parameters of the two layers.

\begin{figure}[t]
\centering
\includegraphics[width=\linewidth]{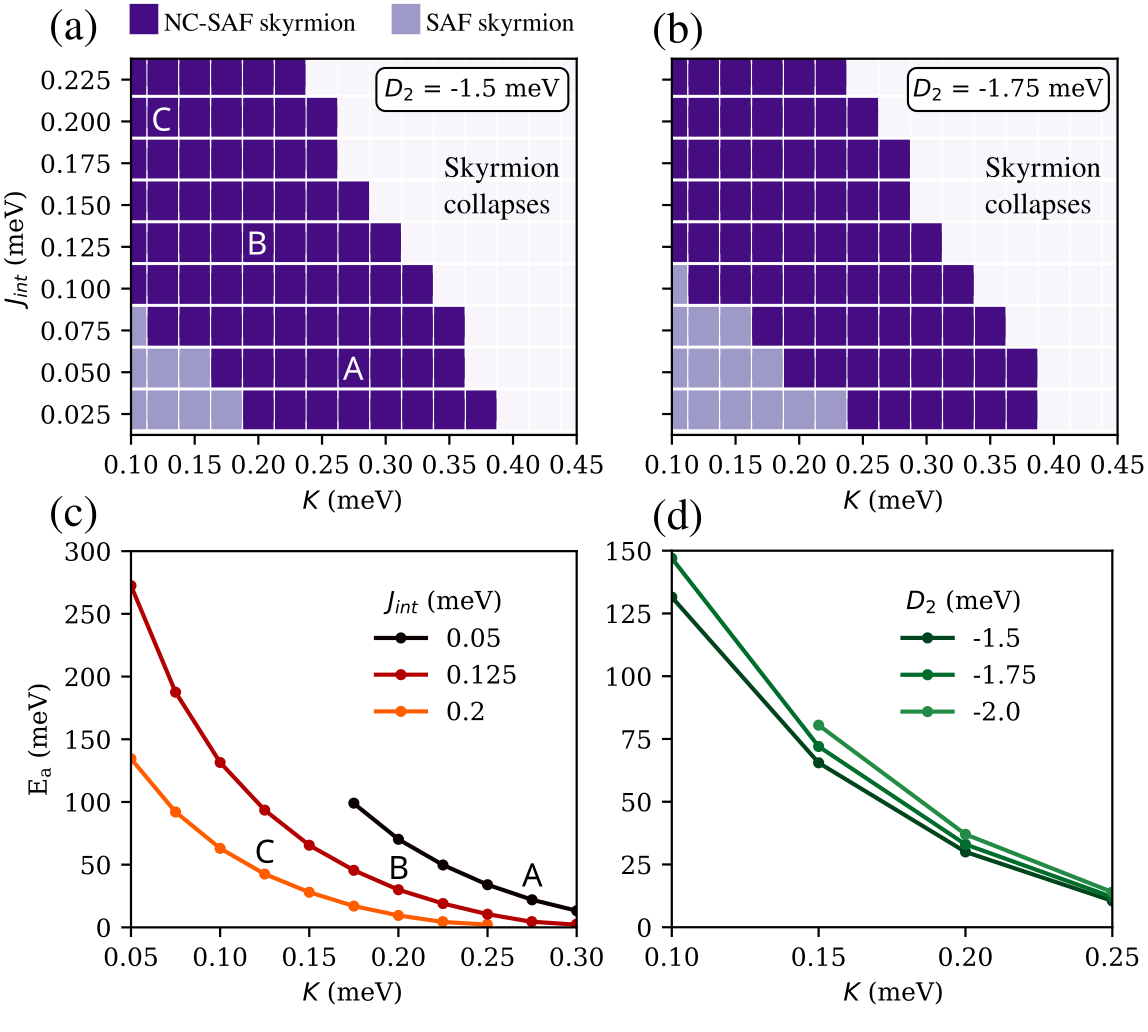}
\caption{ (a,b) Phase diagrams for skyrmion stability obtained from simulations as a function of \( J_\text{int} \) and \( K \), for \( D_1 = 1.5~\text{meV} \) with (a) \( D_2 = -1.5~\text{meV} \) and (b) \( D_2 = -1.75~\text{meV} \). The dark and light purple regions indicate the parameter ranges where NC-SAF and conventional SAF skyrmions are stabilized, respectively. Three points in the diagram shown in (a), marked as A, B, and C, are highlighted for comparison of the activation energies shown in (c). (c) Calculated activation energy profiles as a function of \( K \) for \( D_2 = -D_1 = -1.5~\text{meV} \) and different values of the interlayer exchange \( J_\text{int} \). (d) Calculated activation energy profiles as a function of \( K \) for \( J_\text{int} = 0.125~\text{meV} \) and different values of the DMI strength \( D_2 \).
}
    \label{fig7}
\end{figure}

Note that although enhancing the DMI interaction in one of the layers only slightly influences the stability of the NC-SAF skyrmion (as the collapse occurs through the layer with the weaker DMI), combining this scenario with an applied magnetic field provides an effective way to tune skyrmion stability.

Another key parameter governing the stability of NC-SAF skyrmions is the out-of-plane magnetic anisotropy \( K \). In this configuration, skyrmions are aligned through their domain walls, and \( K \) strongly influences the width of these walls.

Figures~\ref{fig7}(a) and (b) present the phase diagrams for skyrmion stability as a function of \( K \) and \( J_\text{int} \), for DMI ratios \( D_2/D_1 = -1 \) and \( D_2/D_1 < -1 \), respectively. The dark and light purple regions indicate the parameter ranges where NC-SAF and conventional SAF skyrmions are stabilized, respectively. Increasing the DMI ratio slightly reduces the region of stability for NC-SAF skyrmions. This effect arises from the increase in the size of one of the skyrmions, which may cause one skyrmion to fit inside the other, thereby forming a coaxial SAF skyrmion state.

The activation energy profile for the collapse of the NC-SAF skyrmion as a function of \( K \) is shown in Fig.~\ref{fig7}(c) for different values of \( J_\text{int} \). The activation energy rapidly increases as \( K \) decreases. For low values of \( K \), the activation energy can reach hundreds of meV, indicating highly stable NC-SAF skyrmions, comparable to conventional skyrmions. This enhanced stability arises from the interlayer interaction between the domain walls of the skyrmions, which are antiferromagnetically aligned in the NC-SAF state. An increase in the domain wall width strongly enhances the coupling between skyrmions in the two layers. Figure~\ref{fig7}(d) presents the activation energy profiles as a function of \( K \) for \( J_\text{int} = 0.125~\text{meV} \) and different values of DMI strength \( D_2 \). Increasing \( D_2 \) only slightly raises the activation energy. The DMI parameter \( D_2 \) mainly affects the skyrmion size in the second layer, but it has little influence on the shape of the domain wall of the skyrmion, which explains the small variation in activation energy.

\begin{figure}[t]
\centering
\includegraphics[width=\linewidth]{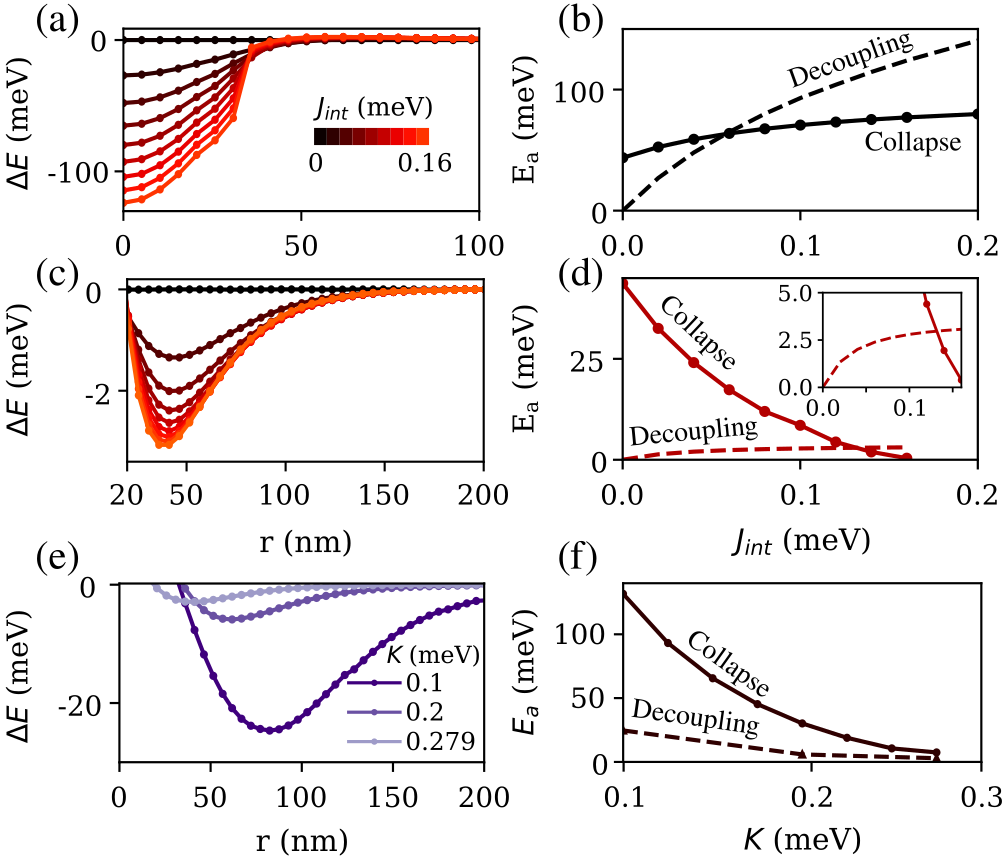}
\caption{
(a,c) Skyrmion--skyrmion interaction energy profiles calculated as a function of the separation distance \( r \) between the skyrmion cores, for different values of \( J_\text{int} \), corresponding to (a) the conventional SAF skyrmion and (c) the NC-SAF skyrmion. (b,d) Corresponding decoupling energies for (b) the conventional SAF skyrmion and (d) the NC-SAF skyrmion. The corresponding collapse energies are also shown for comparison. The inset in (d) presents a zoomed-in view of the energy profiles. Here, \( D_2 = D_1 = 1.5~\text{meV} \) for (a,b), and \( D_2 = -D_1 = -1.5~\text{meV} \) for (c,d). (e) Skyrmion--skyrmion interaction energy profiles for the NC-SAF case with \( J_\text{int} = 0.12~\text{meV} \) and different values of \( K \). (f) Corresponding decoupling energy as a function of \( K \). In (e,f), \( D_2 = -D_1 = -1.5~\text{meV} \) and \( J_\text{int} = 0.12~\text{meV} \).
}
    \label{fig8}
\end{figure}

\subsection{Collapse versus decoupling}

As discussed in the previous sections, the NC-SAF skyrmion exhibits the highest activation energy for its collapse when \( J_\text{int} \to 0 \), showing a monotonic decay of \( E_a \) as \( J_\text{int} \) increases. Considering that a finite positive \( J_\text{int} \) is required to keep the skyrmion pair bound in the bilayer system---thus preventing depairing during its dynamics, as required for applications---the optimal value of \( J_\text{int} \) is the one that simultaneously provides good stability against collapse and strong coupling between skyrmions in the two layers.

In Fig.~\ref{fig8}(a,c), we show the skyrmion--skyrmion interaction energy profiles as a function of the separation distance \( r \) between the skyrmion cores, for different values of \( J_\text{int} \), for both the conventional SAF skyrmion [Fig.~\ref{fig8}(a)] and the NC-SAF skyrmion [Fig.~\ref{fig8}(c)]. To obtain these energy profiles, we fix the magnetic moments at the skyrmion cores in each layer, relax the magnetization, and calculate the total energy as a function of \( r \). For the conventional SAF skyrmion [Fig.~\ref{fig8}(a)], the minimum-energy configuration occurs when the skyrmions are coaxial (\( r = 0 \)), whereas for the NC-SAF skyrmion [Fig.~\ref{fig8}(c)], the energy is minimized at a finite separation between the skyrmion cores, as expected.

The corresponding decoupling energies---defined as the depth of the interaction energy profiles and representing the energy required to separate the skyrmion pair---are shown in Figs.~\ref{fig8}(b,d) for the conventional SAF skyrmion [Fig.~\ref{fig8}(b)] and the NC-SAF skyrmion [Fig.~\ref{fig8}(d)]. The collapse energies are also shown for comparison. Notice that the decoupling energy increases with \( J_\text{int} \), as the skyrmions become more strongly bound. This is in agreement with results presented in literature for similar system~\cite{correia2024stability}.  

Moreover, it becomes evident from Fig.~\ref{fig8}(d) that, for the reference parameters considered in this work (as in Sec.~\ref{secII}~A), the NC-SAF skyrmion not only exhibits a smaller energy barrier for collapse compared to the conventional SAF skyrmion, but is also weakly bound and can be easily depaired, as indicated by the low decoupling energy shown in Fig.~\ref{fig8}(d). \rmm{The optimal stability is achieved at the intersection of the collapse and decoupling energy profiles, occurring at $J_\text{int} \approx 0.13~\text{meV}$ with an activation energy of $E_a \approx 3~\text{meV}$ ($\sim 35~\text{K}$).
}

Based on the results presented in the previous section, and as discussed in the literature~\cite{correia2024stability}, the decoupling energy can be tuned by adjusting the magnetic parameters. In particular, a reduction in the perpendicular anisotropy \( K \) is expected to enhance the interlayer interaction between skyrmions, as it enlarges the skyrmion domain walls that are aligned in the NC-SAF configuration. Figure~\ref{fig8}(e) shows the skyrmion--skyrmion interaction energy profiles for \( J_\text{int} = 0.12~\text{meV} \) and different values of \( K \). As the anisotropy decreases, the decoupling energy (i.e., the depth of the energy profiles in Fig.~\ref{fig8}(e)) increases rapidly, as shown in Fig.~\ref{fig8}(f), where the collapse energy is also shown for comparison. \rmm{For $K=0.1$ meV, the pair-breaking activation energy reaches $E_a=25$ meV ($= 290$ K). Note that increasing the number of atomic layers in the FM layers can further increase the activation energy, thereby increasing the operating temperature range of the noncoaxial skyrmion pair.}

\rmm{Within a particle-like description and assuming strong Gilbert damping $\alpha$, where gyrotropic effects can be neglected, thermally-induced dissociation of the pair is expected to follow an Arrhenius law, with lifetimes given by $\tau=\tau_0e^{E_a/k_BT}$, where $\tau_0$ is the inverse attempt rate~\cite{Zaccone2012}. While a detailed investigation of lifetimes is beyond the scope of this paper, we propose a crude estimate to provide an order-of-magnitude sense. Qualitatively, $\tau_0\sim L^2/\Gamma$, where $L$ is the range of the attractive tail of the skyrmion-skyrmion potential and $\Gamma\simeq k_BT/\alpha\mathcal{D}$ is the skyrmion diffusion constant in the strong damping limit, with $\mathcal{D}$ the dissipation tensor of an isolated skyrmion. Therefore, in this limit
\begin{equation}
  \tau \sim \frac{\alpha\mathcal{D}L^2}{k_BT}e^{E_a/k_BT}
\end{equation}
For $J_\text{int}=0.12$ meV and $K=0.1$ meV, one has $L\sim150$ nm, $\mathcal{D}\sim10^{-14}$ Js/m$^2$, and pair-breaking activation energy $E_a=290$ K, as estimated above. In this case, for temperatures $T=10$ K, 100 K, and 300 K, one gets $\tau\sim 10^6$ s, $10^{-6}$ s, and $10^{-7}$ s, respectively. Therefore, for the chosen system parameters, one can expect very long lifetimes of self-propelling skyrmion pairs under cryogenic temperatures, but short lifetimes near room temperature. Note that lifetimes on the scale of microseconds at room temperature can still be useful for micron-scale applications of self-propelled skyrmions, since typical propulsion speeds are on the order of 1 m/s~\cite{deSouzaSilva2025}.} 

\section{Conclusions}\label{sec.conclusion}

In summary, we have investigated the stability of bilayer skyrmion pairs in synthetic antiferromagnets (SAF) against unbinding and collapse, processes that can compromise device reliability and active skyrmionic functionalities. For the case of homochiral SAF, the preferred configuration for high-speed skyrmionic devices, we obtain that the activation energy for the skyrmion collapse increases monotonically with interlayer coupling and DMI strength $D$, but is a decreasing function of the perpendicular anisotropy constant $K$. The stability can be further enhanced with the help of a symmetry-breaking dc magnetic field, which introduces a mismatch in skyrmion sizes, inducing an increase in the activation energy.  

The situation for non-coaxial skyrmion pairs in homochiral SAF is considerably more complex. First, contrary to intuition, the barrier for the pair collapse decreases monotonically with the interlayer coupling constant $J_\text{int}$. On the other hand, the binding energy increases with $J_\text{int}$. These opposing trends yield an intersection point corresponding to the interlayer coupling that maximizes the overall pair stability. We also found that a suitable combination of DMI ratio and dc magnetic field can further enhance the stability of the pair. Furthermore, the stability can be considerably improved by choosing small perpendicular anisotropy constants, \rmm{pointing to the feasibility of experiments on self-propelled skyrmions over a wide temperature range up to near room-temperature conditions.}

Overall, our results demonstrate that the stability of bilayer SAF skyrmions arises from a delicate competition between collapse and unbinding processes, each controlled by distinct magnetic parameters. This understanding provides concrete guidelines for selecting parameter regimes that enhance robustness in SAF-based racetracks and related skyrmionic technologies.

Finally, although our investigation represents an important first step towards understanding the stability of bilayer skyrmions in SAF materials, many open questions remain. These include the influence of extrinsic material details, such as layer-thickness variations and defects, on the collapse pathways and activation energies. Another line of investigation concerns the role of soft deformation modes excited by spin-polarized currents or oscillating fields. Such questions are particularly relevant for active skyrmionic systems, where the very mechanism of self-propulsion relies on deformation modes excited by ac fields, which might produce significant changes in energy barriers for strong ac amplitudes.

\section*{Acknowledgments}


This work was supported by Coordenação de Aperfeiçoamento de Pessoal de Nível Superior – Brasil (CAPES), Finance Code 001, and Fundação de Amparo à Ciência e Tecnologia do Estado de Pernambuco (FACEPE), Grant Number APQ-1129-1.05/24. INCT project Advanced Quantum Materials (Proc. 408766/2024-7), involving the Brazilian agencies CNPq , FAPESP, and CAPES. CCdSS is funded by Conselho Nacional de Desenvolvimento Científico e Tecnológico – Brasil (CNPq), Grant No{.} 307425/2025-8.

\bibliography{references.bib}
\end{document}